# STUDY OF OXYGEN DIFFUSION AND CLUSTERING IN SILICON USING AN EMPIRICAL INTERATOMIC POTENTIAL

Z. JIANG AND R. A. BROWN
Department of Chemical Engineering, Massachusetts Institute of Technology, Cambridge, MA 02139

## ABSTRACT

The diffusion path and diffusivity of oxygen in crystalline silicon are computed using an empirical interatomic potential which was recently developed [1] for modelling the interactions between oxygen and silicon atoms. The diffusion path is determined by constrained energy minimization, and the diffusivity is computed using jump rate theory. The calculated diffusivity $D=0.025 \ exp(-2.43eV/k_BT) \ cm^2/sec$ is in excellent agreement with experimental data. The same interatomic potential also is used to study the formation of small clusters of oxygen atoms in silicon. The structures of these clusters are found by NPT molecular dynamics simulations, and their free energies are calculated by thermodynamic integration. These free energies are used to predict the temperature dependence of the equilibrium partitioning of oxygen atoms into clusters of different sizes. The calculations show that, for given total oxygen concentration, most oxygen atoms are in clusters at temperature below 1300K, and that the average cluster size increases with decreasing temperature. These results are in qualitative agreement with the effects of thermal annealing on oxygen precipitation in silicon crystals.

## INTRODUCTION

Silicon crystals grown by Czochralski processes contain oxygen as an impurity with a concentration as high as 50 ppm. The oxygen is introduced into the crystal during solidification from the melt because of the dissolution of the quartz crucible used in the Czochralski puller [2,3]. Single oxygen atoms exist in silicon as interstitials that occupy the puckered bond-center sites that bridge two neighboring silicon atoms along the <111> direction [4]. Because the solubility of oxygen in silicon decreases with decreasing temperature, crystal cooling and thermal annealing leads to the precipitation of oxygen clusters that range in size from several nanometers to tens of micrometers. Small clusters of silicon and oxygen containing up to 20 oxygen atoms are electrically active and are referred to as thermal donors. Larger oxygen precipitates, believed to be microphases of silica, play important roles in the internal gettering of metallic impurities in the industrial processing of silicon. Oxygen precipitations also have been found to affect the gate oxide integrity (GOI) characteristics of MOS devices [5].

The oxygen precipitation processes are controlled by the thermodynamics of oxygen clusters and the kinetics of oxygen transport. While there have been many experimental studies of thermal donors, the structural and thermodynamic properties of oxygen clusters remain poorly understood. Experimental measurement [4-11] of the diffusivity of oxygen in silicon, either based on oxygen transport or relaxation of stress-induced-infrared-dichroism method has received much more attention. Although these experiments were performed in different and limited temperature ranges, most data from these experiments can be consistently fit to a single expression of the form $D=0.13 \ exp(-2.53eV/k_BT) \ cm^2/sec$, as pointed out by Mikkelsen [12], who obtained this expression by fitting to data from six independent experiments. This expression is generally believed to be the intrinsic diffusion constant involving oxygen jumping from a bond-center site to one of the six nearest bond-center sites.

Theoretical investigation of oxygen diffusion and oxygen clusters in silicon is difficult because of the large diffusion barrier and the complicated nature of the mixed bonding between oxygen and silicon atoms. Several researchers [13-17] have attempted to calculate the energy barrier of oxygen diffusion and the structures of small oxygen clusters using local den-sity functional theories or self-consistent cluster calculations. These efforts have yielded diffusion barriers ranging from 1.2eV to 2.5eV. All these calculations assume that the saddle point configuration for diffusion is in a (110) plane and at the midway between the two bond-center sites. No calculations of the prefactor of the diffusion constant have been reported. A variety of stable and metastable structures of small oxygen clusters containing up to three oxygen atoms also have been reported [13-17].

We have studied oxygen diffusivity and oxygen clusters in silicon using an empirical potential that has recently been developed for oxygen-silicon interactions [1]. The diffusion path and diffusion barrier are determined [18] by a series of energy minimization calculations performed with the constraint of a constant cone angle θ between the O-Si bond and the axis connecting the two silicon atoms bonded to the oxygen atom in the initial equilibrium configuration; the angle θ is shown in Fig.1. The diffusivity is calculated using jump rate theory [19,20]. The structures of small oxygen clusters are found by NPT molecular dynamics simulations using the same interatomic potential. The free energies of these clusters are calculated by thermodynamic integration and used to predict the temperature dependence of the equilibrium partitioning of oxygen atoms into clusters of different sizes.

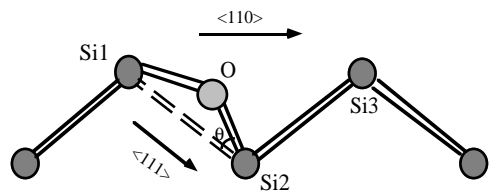

FIG. 1. Schematic diagram showing the constraint angle θ.

In the remainder of this paper, we first present our computational methods, followed by a summary of the results of our calculations of oxygen diffusivity in silicon, which have been reported in an earlier publication [18]. The results of the calculations of small oxygen clusters and oxygen partitioning into these clusters are then described.

## COMPUTATIONAL METHODS

Empirical interatomic potential

The empirical interatomic potential for the OSi system is constructed [1] using the Stillinger-Weber (SW) silicon potential [21] and the silica potential of van Beest *et al.* [22]. Three important new components are introduced to describe the charge-transfer and mixed bonding between oxygen and silicon atoms. These ingredients include (a) a charge transfer function, (b) a bond-softening function, and (c) an ionization energy. The parameters are fitted to the structure, vibrational frequency, and formation energy of an oxygen interstitial in silicon. The formulation of this composite interatomic potential, the parameter values, and the details of the construction and parameter fitting process are reported in [1], where the potential is also used to study the structural, energetic and vibrational properties of interstitial-oxygen and vacancy-oxygen clusters, yielding results in good agreement with available experiments on vacancy-oxygen structure.

Constrained energy minimization and jump rate theory

The diffusion path and the diffusion energy barrier are determined from a series of energy minimization calculations which are performed with the constraint of a constant cone angle θ, as shown in Fig. 1. This constraint allows the oxygen atom to move out of the (110) plane, and only reduces one degree of freedom of the system. The constraint is implemented using the augmented Lagrangian method [23] in which the generalized energy function

$$E_c\big(\{r_i\}, \lambda\big) = E\big(\{r_i\}\big) + \lambda g(r_{20}, r_{21}) + \frac{1}{2}w g^2(r_{20}, r_{21}) \qquad (1)$$

is minimized by the steepest descent method. In Eq.(1), $E(\{\mathbf{r}_i\})$ is the energy of the unconstrained system, $\lambda$ is the Lagrangian multiplier,

$$g(\mathbf{r}_{20}, \mathbf{r}_{21}) = \frac{\mathbf{r}_{20} \cdot \mathbf{r}_{21}}{r_{20} r_{21}} - \cos\theta \tag{2}$$

is the constraint function, where $\mathbf{r}_{20}$ is the vector from Si2 to O and $\mathbf{r}_{21}$ is the vector from Si2 to Si1, as shown in Fig.1. The value of $\theta$ is increased gradually from 0 and the converged configuration for each $\theta$ is used as the initial configuration for the steepest-descent minimization at the next larger value of $\theta$.

The diffusion constant is given by the jump rate theory [19,20] as an Arrhenius expression $D = D_0 \exp(-\Delta E/k_B T)$, where $\Delta E$ is the diffusion energy barrier, and the prefactor $D_0$ is determined by $D_0 = g l^2 \nu/(2d)$, where $d$ the dimensionality of space, $l$ is the elementary jump length, and $g$ is the number of equivalent diffusion paths [19]. Because each bond-center site has six nearest neighbors and six fold degeneracy [4,19,24], $g=36$. The attempt frequency $\nu$ is given by [20]

$$\nu = \left(\prod_{i=1}^{3n} \nu_i^e\right) \Big/ \left(\prod_{i=1}^{3n-1} \nu_i^s\right) \tag{3}$$

where $\{\nu_i^e\}$ are $3n$ frequencies of the n-atom system in the equilibrium state, and $\{\nu_i^s\}$ are the $3n-1$ frequencies of the system at the saddle point configuration along the diffusion path. The frequencies $\{\nu_i^e\}$ and $\{\nu_i^s\}$ are calculated by direct diagonalization of the Hessian matrix of the system at the equilibrium and saddle point configurations, respectively.

NPT molecular dynamics and thermodynamic integration

The equilibrium structures of oxygen clusters are determined by NPT molecular dynamics simulations with zero pressure. The algorithm of Andersen [25] is used to keep the pressure constant and the integration of the equations of motion is performed numerically using the fifth-order predictor-corrector method [26]. Constant temperature is achieved by scaling the momentum. Simulations are performed with a time step of 0.1536 fs. The simulation cell has periodic boundaries and contains 216 silicon atoms and N oxygen atoms.

The free energy $G(T)$ is calculated using thermodynamic integration, i.e,

$$G(T) = T\left(\frac{G_0}{T_0} - \int_{T_0}^{T} \frac{H(T)}{T^2} dT\right) \tag{4}$$

where $G_0$ is the free energy calculated using the harmonic approximation [1,27] at a reference temperature $T_0 = 500K$, and $H(T)$ is the enthalpy of the system, as computed by molecular dynamics simulations at temperature $T$.

**RESULTS**

Oxygen diffusion path and diffusivity

The computed energy of the system with one oxygen atom is shown in Fig. 2 as a function of $\theta$. The saddle point configuration, with an energy barrier of $\Delta E = 2.43\ eV$, is found to occur at $\theta=78^o$. The diffusing oxygen atom reaches the top of the energy barrier at a point beyond the midpoint between the two bond-center sites, which corresponds to $\theta=55^o$. Atomic configurations [18] show that the diffusion path is primarily in the (110) plane. In the diffusion process, only the three silicon atoms involved in the O-Si bonds are significantly displaced from their lattice positions. As the oxygen atom moves from the original equilibrium site, one of the Si-O bonds is stretched, but is not broken until the oxygen moves through the saddle point at $\theta=78^o$. Beyond the saddle point, the stretched bond is broken and a new Si-O bond is formed with a silicon atom in the other <111> direction, thus forming a new bridging configuration for the oxygen. This oxygen migration process is consistent with the numerical experiments of kicked oxygen performed by Needels *et al.* [13].

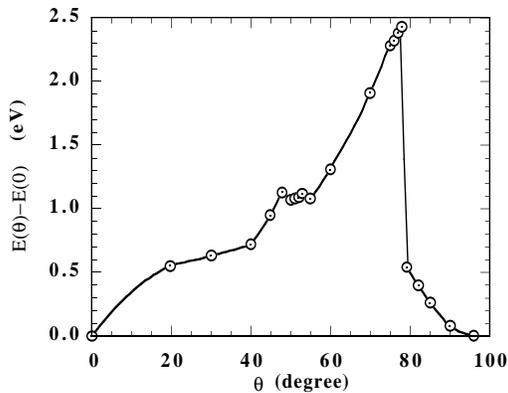 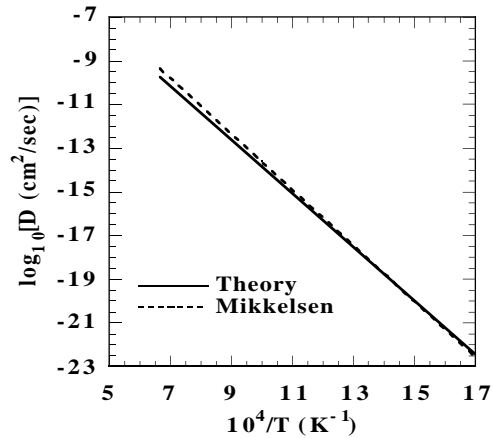

FIG. 2. The computed energy as a function of θ for the system with one oxygen atom. The saddle point configuration occurs at θ=78°.

FIG. 3. Comparison of the calculated expression for the oxygen diffusivity with Mikkelsen's best fit to experimental data[10].

The jump frequency is computed from Eq.(3) to be *v=11.5THz*. In the equilibrium state, the distance between two nearest neighbor bond-center sites is *l =1.91Å*. The diffusivity is predicted as

$$D = 0.025 \exp(-2.43 eV/k_B T) \, cm^2/sec \qquad (5)$$

This calculated result is plotted in Fig. 3 together with Mikkelsen's best fit [12] to six sets of independent experimental data, $D=0.13 \exp(-2.53eV/k_BT) \, cm^2/sec$. The agreement is excellent although the prefactors and energy barriers are different. The difference is well within range of experimental error in the data. Indeed, Newman *et al.* [11] reported an Arrhenius fit of the form $D=0.02 \exp(-2.42eV/k_BT) \, cm^2/sec$ to data from an experiment relating to the kinetics of oxygen precipitation. His expression is virtually the same as our calculated result.

Oxygen clusters and partitions

The structures of the $O_N$ (N=2, 3, 4) clusters are obtained as shown in Fig. 4. The $O_2$ and $O_3$ clusters are both in a (110) plane. They are similar to those found by Needels *et al.* [13], but the two oxygen atoms in $O_2$ are not bonded to a common silicon atom. The $O_4$ cluster has a tetrahedral structure. The formation energies per oxygen atoms are *0.47eV*, *0.80eV* and *0.13eV* for N=2, 3 and 4, respectively. Because the formation energy for single oxygen interstitial $O_1$ is *1.06eV* [1], the $O_2$ structure has a binding energy of *1.18eV*. The $O_3$ structure has an energy *0.28eV* higher than the energy of an ($O_1+O_2$) configuration, *it is therefore only metastable*. The $O_4$ cluster is stable with respect to both the $4O_1$ and $2O_2$ configurations.

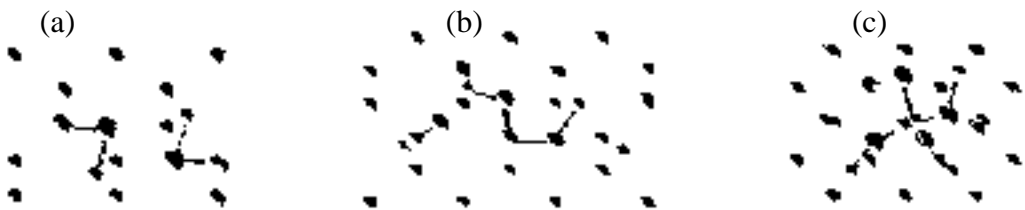

FIG. 4. The structures of small oxygen clusters: (a) $O_2$ viewed at the [110] direction; (b) $O_3$ viewed at the [110] direction; (c) $O_4$ viewed at the [100] direction. The oxygen atoms are slightly larger, and connected by two bonds.

The free energies of $O_1$, $O_2$, and $O_4$ are calculated using the thermodynamic integration, Eq.(4). The results are shown in Fig. 5. Assuming that the total concentration of oxygen is 20 ppm and that oxygen only exists in the forms of either $O_1$, $O_2$, or $O_4$, we have calculated the equilibrium partitioning of oxygen atoms. Fig. 6 displays the fractions of oxygen atoms in the forms of $O_1$, $O_2$ and $O_4$ as a function of temperature. As temperature decreases, more and more oxygen atoms are partitioned into larger clusters. Below 1300K, less than 3% of oxygen atoms exist in the form of single interstitials. This is consistent with the experimental finding that pre-heating at a temperature higher than 1000C$^o$ is effective in dissolving oxygen clusters and generating single oxygen interstitials [28].

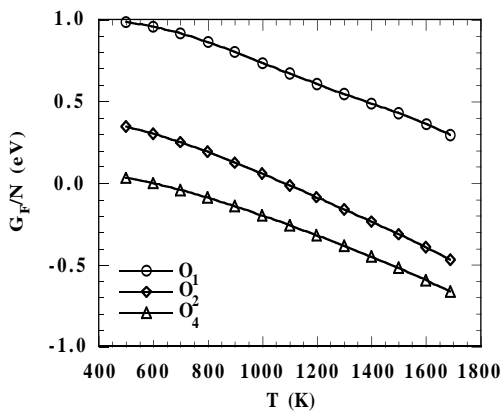

FIG. 5. The formation free energy per oxygen atom for oxygen clusters of size 1, 2 and 4.

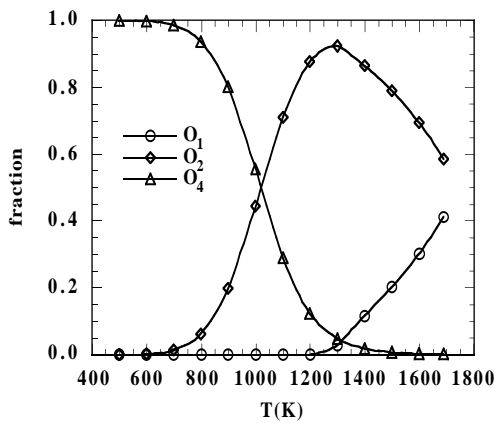

FIG. 6. The fraction of oxygen atoms in the form of $O_1$, $O_2$, and $O_4$ for a given total concentration of oxygen of 20ppm.

## CONCLUDING REMARKS

We have presented theoretical calculations of oxygen diffusion and oxygen clustering in crystalline silicon using an empirical potential constructed to describe the complicated charge-transfer and bond-formation processes between silicon and oxygen atoms. We found the oxygen jumps from a bond-center site to another bond-center site along a path in the (110) plane. Because the oxygen atom has to break a Si-O bond before it is able to form a new bridging configuration with the other silicon atom, the saddle point is predicted to be farther than the midpoint between the two bond-center sites. The diffusion coefficient predicted using this saddle point configuration and jump rate theory is in excellent agreement with experiments. We also found that the $O_3$ zigzag structure along a [110] direction is energetically unstable with respect to the ($O_1+O_2$) configuration, which suggests that clusters of even number of oxygen atoms are favored in equilibrium. This is particularly interesting in the context of proposed possible role of $O_2$ in the thermal donor formation processes [29]. Our calculations of oxygen equilibrium partitioning into clusters of different sizes show that, for a given oxygen concentration, most oxygen atoms are partitioned into clusters at temperature below 1300K, and that the average cluster size increases with decreasing temperature. These results are in agreement with the general effects of thermal annealing on oxygen precipitation in silicon crystals.

## ACKNOWLEDGMENT

We thank MEMC Corporation, Sematech, Shin-Etzu Handotai Company, and Wacker Chemitronic for financial support of this research.